# The implications of large-scale containment policies on global maritime trade during the COVID-19 pandemic

J.Verschuur[1,*], E.E. Koks[1,2] and J.W. Hall[1]

[1]Environmental Change Institute, University of Oxford, Oxford, United Kingdom

[2]Institute for Environmental Studies, Vrije Universiteit Amsterdam, Amsterdam, Netherlands

[*]Corresponding author

Email: jasper.verschuur@keble.ox.ac.uk (JV)

# Abstract

**The implementation of large-scale containment measures by governments to contain the spread of the COVID-19 virus has resulted in a large supply and demand shock throughout the global economy. Here, we use empirical vessel tracking data and a newly developed algorithm to estimate the global maritime trade losses during the first eight months of the pandemic. Our results show widespread trade losses on a port level with the largest absolute losses found for ports in China, the Middle-East and Western Europe, associated with the collapse of specific supply-chains (e.g. oil, vehicle manufacturing). In total, we estimate that global maritime trade reduced by -7.0% to -9.6% during the first eight months of 2020, which is equal to around 206-286 million tonnes in volume losses and up to 225-412 billion USD in value losses. The fishery, mining and quarrying, electrical equipment and machinery manufacturing, and transport equipment manufacturing sectors are hit hardest, with losses up to 11.8%. Moreover, we find a large geographical disparity in losses, with some small islands developing states and low-income economies suffering the largest relative trade losses. We find a clear negative impact of COVID-19 related business and public transport closures on country-wide exports. Overall, we show how real-time indicators of economic activity can support governments and international organisations in economic recovery efforts and allocate funds to the hardest hit economies and sectors.**

# Introduction

The emergence and spread of COVID-19, caused by the severe acute respiratory syndrome coronavirus 2 (SARS-CoV-2), has forced countries worldwide to implement large-scale containment measures to reduce the spread of the virus [1–4]. These measures, which include among others international travel restrictions, business closures, prohibition of large-scale private and public gatherings, and mandatory quarantines, have shown to effectively reduce the rate of transmission of the virus [1,3,5]. As a consequence, however, such policies have had large economic repercussions, both in terms of domestic industry output and international trade, due to diminishing production and reduced demand for some products. The resulting demand and supply shocks has further cascaded through complex supply-chains networks, causing spill-over effects to downstream and upstream suppliers, domestically and internationally [6–8].

Previous research has intended to estimate the global macroeconomic impacts of enforcement measures worldwide using model-based approaches, in particular input-output (I-O) analysis and computable general equilibrium (CGE) models [6–9]. For instance, Inoue and Todo [8] estimate that an one month lockdown of Tokyo may lead to a 5.2% contraction of the annual gross domestic production (GDP) of Japan. Lenzen *et al.* [7] estimate the global macroeconomic losses to be around 4.2% of GDP, whereas Guan *et al.* [6] estimate the losses in global value-added to be between 25-40.3%, depending on the modelling scenario adopted. In addition, the World Trade Organisation (WTO) projects that global international trade will contract by 13-32% in 2020 under a number of scenarios of disruptions to economic activity [10]. However, the accuracy of these model-based estimates critically hinge on setting realistic scenarios of the economic impacts and understanding how supply-chain dynamics disrupt or buffer shocks. This includes making assumptions how enforcement policies impact economies and predicting how countries recover from their economic downturn. Given the large heterogeneity in containment policies implemented, in terms of severity, duration and enforcements [1,11], alongside the complexity of shock propagation through interconnected networks [12,13], the extent to which such model-based approaches can predict the nature and scale of the economic impact currently remains to be seen. Moreover, there is a lack of real-time global observational data to validate and steer model projections, which may cause ill-informed decision-making or misallocation of funds [5]. For instance, trade statistics, published by national statistical agencies, are usually produced with some delay and are more comprehensive for OECD countries. Hence, alternative data sources are needed to provide real-time indicators of economic activity in order to fill existing data gaps and identify trends. For instance, a report by UNCTAD [14] estimated that in the first half of 2020, global port calls have declined by 8.7%, with the largest drop estimated to be in Europe, Australia and Oceania.

Here, we use a novel dataset derived from empirical vessel tracking data in combination with a newly developed algorithm to estimates the changes in international maritime trade flows during the outbreak of the COVID-19 pandemic. Maritime trade flows cover around 80% of the world's trade in terms of volume [15], and can hence be used as a first-order indicator of the status of economic activity in a country. The newly derived dataset has a high spatial (166 countries) and temporal (daily) resolution, which helps to track the impacts of the COVID-19 outbreak the global economy, and the geographical disparity of impacts. To assess the impacts, we first quantify the total trade losses and gains for ports and countries for the period January-August by comparing the 2020 data to the same period in 2019. Second, we provide a sector-level disaggregation to show how certain sectors have been disproportionally affected by supply and demand changes. Third, we use a panel regression model with fixed effects to estimate the impacts of specific containment policies on exports by making use of the heterogeneity in the diversity, timing and severity of containment measures across countries. Overall, we provide evidence how the impacts to global trade are complex and dependent on the trade-dependencies, sector-composition and policies implemented. The results provide timely information about which economies are hit hardest, thereby helping decision-makers effectively target and prioritize international aid and economic stimulus.

# Method

## Data and trade estimation

We derive estimates of port-level trade flows (imports and exports) for 1153 ports across 166 countries worldwide using the geospatial location and attributes of maritime vessels (from January 2019 - August 2020). To do this, we use Automatic Identification System (AIS) data, which provides detailed data on the location, speed, direction and vessel characteristics of all trade-carrying vessels with an AIS transponder (that send information to terrestrial or satellite receivers every few seconds-minutes) [16]. This data was obtained through a partnership with the UN Global Platform AIS Task Team initiative (https://unstats.un.org/wiki/display/AIS/), which aims to develop algorithms and methodologies to make AIS data useful for a variety of fields and applications (traffic, economic trade, fisheries, $CO_2$ emissions).

We develop an algorithm (S1 Appendix) that estimates the trade flows based on the ingoing and outgoing movements of maritime vessels (~3.2 million port calls across 100,000 unique vessels) and their characteristics (e.g. dimensions, utilisation rate, vessel type), going on to disaggregate these trade flows into specific sectors (11 sector classification adopted here). We end up with daily sector-specific trade flow estimates on a port-level, which we aggregate to a country-scale to perform the country-wide impact analysis. This new algorithm significantly advances previous work [17–19] by providing a global scale analysis and being able to provide a sector decomposition.

We validate the results (S1 Appendix) by comparing the derived trade estimates to detailed port-level trade data obtained for five countries (Japan, United Kingdom, United States, New Zealand, Brazil). Moreover, we compare our estimates to country-wide maritime trade flows obtained from UN Comtrade [20] mode of transport data for 27 countries.

## Econometric model

The variation in trade losses among countries are driven by the differences in containment measures introduced by countries (in terms of timing, duration, and severity) [5], and due to supply shortages to domestic supply-chains and demand reductions in trade-dependent economies [6]. Large-scale containment measures can directly negatively influence industry output by affecting business operations (e.g. workplace closure, mobility restrictions), or indirectly positively affect industry output through effectively containing the virus outbreak and thereby allowing industrial production processes and transportation of goods to continue. To study the implications of large-scale containment measures on exports (which we use as a proxy of industrial output), we match our daily, country-wide, estimates with data from the Oxford COVID-19 Government Response Tracker (OxCGRT) [21]. Within OxCGRT, data is collected on the implementation and stringency of containment measures across 160 countries. We utilise reduced-form econometric techniques [22] to estimate the effect of different containment policies on exports across a balanced sample of 122 countries. We express export change as the percentage change in detrended exports in 2020 compared to 2019 (S2 Appendix), which therefore controls for potential seasonality and trends in the data. The time series is first smoothed using a 10-day moving average in order to remove the daily noise and better capture the underlying signal. We further control (see S2 Appendix for discussion) for several factors on a daily-scale, including the number of confirmed cases as a fraction of the population, reduction in demand in trade-dependent countries, the potential reduction in exports due to supply-shortages (through imports that are used for exports, see Hummels et al. [23]), and other endogenous factors that are likely to be serially correlated with exports (which we control for by adding a lag of the export change).

From the OxCGRT [21], we obtained information on nine measures that potentially affect business operations: C1 - School closing; C2 - Workplace closing; C3 - cancel public events; C4 - Restrictions on gatherings; C5 - Close public transport; C6 - Stay at home; C7 - Restrictions on internal movement; C8 - International travel controls; H2 - Testing policy. We scale the severity of the policies on a scale between 0 and 1, thereby assuming a linear relationship between maritime exports and the severity of policies. Moreover, we create a composite stringency index (Stringency) of all policies (C1-C8) by adding all individual policies together and rescaling the index between 0 and 1. We performed multiple robustness checks to evaluate the influence of different model specifications (see S2 Appendix).

# Results

## Model validation

We find a good fit between the values predicted by our algorithm and the reported trade flows on a port-level (correlation coefficient between 0.52-0.96) and a country-level (correlation coefficient between 0.79-0.98), with a general overestimation for smaller ports, and ports and countries with large trade imbalances (e.g. small islands). For the external validation data, we find correlation coefficients of 0.84-0.86 for the aggregated trade data and 0.73-0.73 for the sector-specific trade data (on a country level). Again, smaller trade flows are harder to predict. The accuracy of the method is also found to be dependent on the coverage of information in the AIS data (some attributes are manually put in), especially information on the vessel draft, which is less frequently reported in developing countries.

## Port-level trade flows

In the first eight months of 2020, the number of port calls across all ports reduced by 4.4% compared to same months in 2019. Fig 1a shows the average change in total trade (imports + exports) in terms of volume (in million tonnes, MT) over the months January-August. The vast majority of ports have experienced a decline in total trade, although a number of ports in Brazil, the Gulf of Mexico region, the Middle-East, Australia, and parts of South-Korea and the Philippines have seen an increase in trade in 2020 relative to 2019. The top 20 port with the largest changes in volume in terms of total trade, imports and exports are included in Table 1. The ports with the largest absolute changes in volume are the ports of Ningbo (China, -68.5 MT), Rotterdam (Netherlands, -43.2 MT), Shanghai (China, -32.5 MT), Wuhan (China, -21.6 MT) and Tubarao (Brazil, -20.7 MT). The largest changes in imports are found for the ports of Ningbo (China, -43.5 MT), Rotterdam (Netherlands, -40.1 MT), Shanghai (-22.4 MT), Zhoushan (China, -22.4 MT) and Amsterdam (Netherlands, -12.2 MT). These ports, and the other ports in the list, function as major gateway ports for a country to import final products (New York-New Jersey, Rotterdam), or are essential for specific supply-chains, such as textiles and electronics manufacturing (Shanghai, Ningbo, Zhoushan), steel and paper manufacturing (Ghent, Amsterdam, Rizhou), car manufacturing (Yokohama) and raw materials (coal imports for Krishnapatnam). The largest export changes are found for the ports of Ningbo (China, -25.0 MT), Tubarao (Brazil, -17.1 MT), Novorossiysk (Russia, -11.5 MT), Wuhan (China, -10.8 MT), Beaumont (USA, -10.6 MT) and Dampier (Australia, -10.2 MT). These ports, and the other top 20 ports with largest export losses, are all important export ports for global supply-chains, including the exports of iron ore (Dampier, Tubarao), coal (Haypoint), oil and refined petroleum products (Puerto Bolivar, Fujairah, Beaumont and Novorossiysk) and manufacturing products (Ningbo, Wuhan and Shanghai).

Fig 1b-m show the changes in total trade per month for all ports and the cumulative changes in trade over latitude and longitude. In January, losses are predominantly pronounced in China that extended their Lunar New Year holiday [24], among other measures, resulting in output losses to the Chinese industry. This resulted in a direct demand shock, in particular for the export of raw materials (e.g. iron ore, copper, nickel) that China predominantly imports [25]. This can be observed from the large negative losses found in the large export ports of Brazil. In February, ports in Europe experienced their first drop in imports (blue line top plot), while export losses are still concentrated in Asia. This import drop in Europe coincided with the transit time from China to Europe, which is around three weeks. The export drop is, alongside Brazil,

also visible in the main iron ore exporting ports in Australia (Port Hedland and Port Walcott) and South Africa (Port of Richards Bay) that both supply iron ore to the Chinese industry. In March, exports temporally recovered, while imports dropped in many parts of the world, mainly to due to initiation of lockdowns in economies outside Asia. In particular, India, Malaysia, Singapore, USA West Coast and Mexico have seen a large drop in trade in this month. In April, trade partly recovers in the Northern Hemisphere, while in May the second drop in global trade hits the global economy, as a widespread reduction in demand and supply ripple through the economy. Losses are again pronounced in China and Western Europe, leading to the lowest total import and exports changes on a global scale. In June, July and August, a partial recovery is visible for some ports, while the Middle-East, Eastern Australia, Japan and Western Europe (in particular Belgium and the Netherlands) show large losses. For the Middle-Eastern countries, the collapse of the oil market has contributed to the large trade losses (which are predominantly exports losses). In August, signs of recovery (especially imports) are visible for the Philippines, India, South Africa, Brazil and Argentina, and parts of the Mediterranean, while other countries are still experiencing large losses.

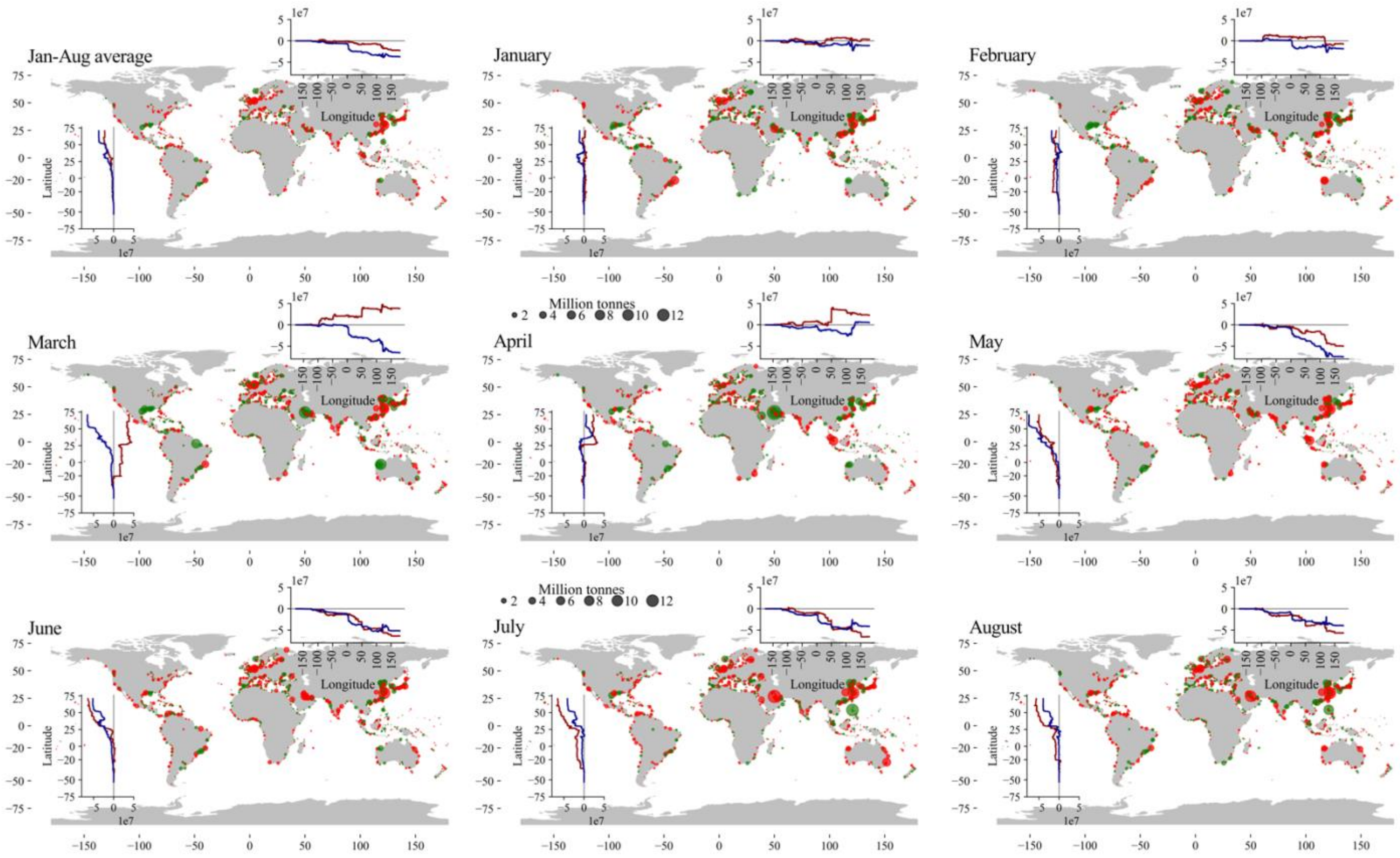

**Fig 1. Port-level trade losses over time.** The geographical location and magnitude of trade losses for Jan-Aug 2020 compared to 2019, including the average over the eight months and the losses per month. Green = positive change, red = negative change. The subplots show the cumulative change over latitude and longitude for imports (dark blue) and exports (dark red).

**Table 1. Largest absolute trade losses on a port-level.** The top 20 total trade, imports and export losses on a port-level expressed in million tonnes (MT). The losses cover the period Jan-Aug 2020 compared to Jan-Aug 2019.

| | **Total trade** | | | **Imports** | | | **Exports** | | |
|---|---|---|---|---|---|---|---|---|---|
| Rank | Port | iso3 | Change (MT) | Port | iso3 | Change (MT) | Port | iso3 | Change (MT) |
| 1 | Ningbo | CHN | -68.5 | Ningbo | CHN | -43.5 | Ningbo | CHN | -25.0 |
| 2 | Rotterdam | NLD | -43.2 | Rotterdam | NLD | -40.1 | Tubarao | BRA | -17.1 |
| 3 | Shanghai | CHN | -32.5 | Shanghai | CHN | -22.4 | Novorossiysk | RUS | -11.5 |
| 4 | Wuhan | CHN | -21.6 | Zhoushan | CHN | -13.8 | Wuhan | CHN | -10.9 |
| 5 | Tubarao | BRA | -20.7 | Amsterdam | NLD | -12.2 | Beaumont | USA | -10.6 |
| 6 | Zhoushan | CHN | -18.8 | Rizhao | CHN | -11.3 | Dampier | AUS | -10.2 |
| 7 | Amsterdam | NLD | -17.4 | Wuhan | CHN | -10.7 | Shanghai | CHN | -10.1 |
| 8 | Shekou | CHN | -14.2 | Mina Al Ahmadi | KWT | -9.5 | Haypoint | AUS | -9.2 |
| 9 | Hong Kong | HKG | -12.3 | Vlissingen | NLD | -8.5 | Lumut | MYS | -7.6 |
| 10 | Vlissingen | NLD | -12.2 | Zhanjiang | CHN | -7.6 | Shekou | CHN | -7.5 |
| 11 | Singapore | SGP | -12.1 | Umm Said | QAT | -7.4 | Tianjin | CHN | -7.3 |
| 12 | Rizhao | CHN | -11.7 | Yokohama | JPN | -7.3 | Fujairah | ARE | -7.2 |
| 13 | Novorossiysk | RUS | -11.7 | Ghent | BEL | -7.1 | Tangshan | CHN | -6.3 |
| 14 | Lumut | MYS | -11.6 | Singapore | SGP | -6.8 | Xiamen | CHN | -6.1 |
| 15 | Dampier | AUS | -10.8 | Hong Kong | HKG | -6.7 | Itaqui | BRA | -5.8 |
| 16 | Yokohama | JPN | -9.9 | Shekou | CHN | -6.7 | Bohai Bay | CHN | -5.7 |
| 17 | Haypoint | AUS | -9.7 | Krishnapatnam | IND | -6.7 | Puerto Bolivar | COL | -5.7 |
| 18 | Beaumont | USA | -9.5 | Magdalla | IND | -6.6 | Hong Kong | HKG | -5.6 |
| 19 | Ghent | BEL | -9.4 | Port of Le Havre | FRA | -6.4 | Primorsk | RUS | -5.5 |
| 20 | Zhanjiang | CHN | -9.1 | New York-New Jersey | USA | -6.0 | Richards Bay | ZAF | -5.4 |

## Geographical disparity

Fig 2a-b show the country-aggregated relative changes in imports and exports, with the top 20 largest negative (relative) changes included in Table 2 and largest negative absolute changes in S1 Table. The top 20 largest total trade losses range between 17-36%. The largest percentage change in imports are associated with small economies such as Turks and Caicos Islands, the Caribbean Netherlands, Bahrain, Anguilla, Federated States of Micronesia and Madagascar (all between 28-37% reduction). Most of the countries with the largest import losses are Small Island Developing States, which are characterised by having large import-dependencies due to their small domestic economies, being reliant on maritime trade flows for trade, and importing large amount of goods to support the tourism sector that constitutes a large share of the country's GDP [26]. With the tourist industry collapsing due to the COVID-19 outbreak [27], the imports are expected to drop further significantly, explaining the widespread reductions observed. Other countries, like a number of countries in Africa, Myanmar, Oman, Philippines, the Baltic States and Sweden have increased their imports, likely due to the increased need for food and medical supplies in developing nations or increased household consumption in some developed countries. In terms of exports, the largest relative losses are found for Libya, New Caledonia, Guinea-Bissau, Northern Mariana Islands, Cape Verde and Sudan (all between 50-78% reduction). These countries include many raw materials exporting countries that have suffered from the demand shock across the world, in particular through trade dependencies with Europe, China and the United States [28]. Moreover, many low income countries had to pro-actively lockdown economies to protect their health care system, or are engaged in economic activities that are less able to be done remotely [28,29]. Some countries have increased their exports, such as India, Myanmar, Vietnam and Philippines,

potentially because of production shifts of manufacturing goods to these countries when factory shut down in China [30]. Moreover, exports grew in Argentina, mainly due to booming exports of food products (e.g. soybeans, beef) to the United States and China [31], and in Tanzania, which increased its exports of gold and food (e.g. nuts) and textile products (e.g. cotton) [32].

Using the World Bank income classification, we test whether high and upper middle income countries have experienced more severe impacts than low and lower middle income countries. Without excluding outliers from the data, we find a significant difference (two-sided t-test with $p>0.05$) between both income groups for exports and imports, with high and upper middle income countries having higher export losses. Hence, although the high and upper middle income countries have higher mean export losses, the most extreme export losses and gains are found for low and lower middle income countries.

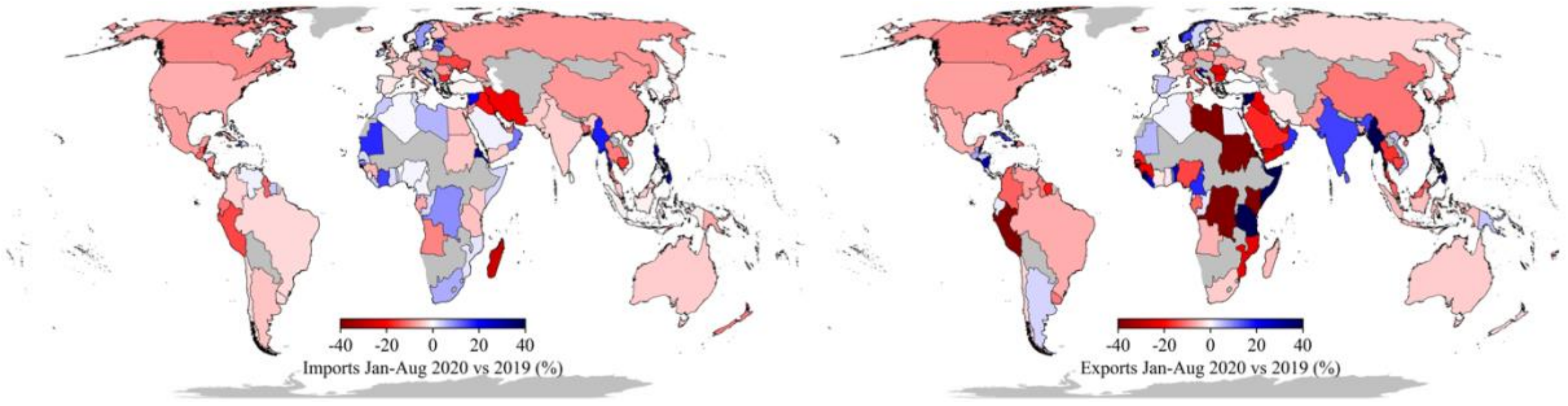

**Fig 2. Country-level relative trade losses.** The relative trade losses for Jan-Aug 2020 compared to 2019 expressed in percentage change. Grey countries indicate no data available.

**Table 2. Largest relative trade losses on a country-level.** The total trade, imports and export losses on a country-level expressed in million tonnes (MT). The losses cover the period Jan-Aug 2020 compared to Jan-Aug 2019.

| | **Total trade** | | **Imports** | | **Exports** | |
|---|---|---|---|---|---|---|
| Rank | Country | Change (%) | Country | Change (%) | Country | Change (%) |
| 1 | Anguilla | -35.6 | Turks and Caicos Islands | -36.9 | Libya | -77.8 |
| 2 | Libya | -34.3 | Bonaire, Saint Eustatius and Saba | -35.9 | New Caledonia | -64.9 |
| 3 | Federated States of Micronesia | -33.5 | Bahrain | -31.3 | Guinea-Bissau | -55.6 |
| 4 | Cape Verde | -30.6 | Anguilla | -30.7 | Northern Mariana Islands | -54.9 |
| 5 | Peru | -28.3 | Federated States of Micronesia | -29.8 | Cape Verde | -53.7 |
| 6 | Bonaire, Saint Eustatius and Saba | -26.8 | Madagascar | -28.4 | Sudan | -49.4 |
| 7 | Malta | -26.2 | Timor-Leste | -26.0 | Montenegro | -45.1 |
| 8 | Eritrea | -26.1 | Malta | -25.7 | Eritrea | -44.6 |
| 9 | Madagascar | -25.1 | Grenada | -22.4 | Dem. Republic Congo | -44.3 |
| 10 | Montenegro | -24.9 | Belize | -22.3 | Vanuatu | -40.0 |
| 11 | Turks and Caicos Islands | -24.8 | Iran | -21.8 | Kenya | -39.6 |
| 12 | Vanuatu | -24.4 | Seychelles | -21.5 | Peru | -39.2 |
| 13 | Seychelles | -23.7 | French Polynesia | -21.1 | Federated States of Micronesia | -38.4 |
| 14 | Timor-Leste | -23.3 | Aruba | -19.8 | American Samoa | -34.9 |
| 15 | Northern Mariana Islands | -22.2 | Vanuatu | -19.4 | Albania | -32.6 |
| 16 | French Polynesia | -21.1 | Iraq | -19.3 | Seychelles | -28.0 |
| 17 | Iraq | -20.1 | Kuwait | -19.0 | Malta | -27.2 |
| 18 | New Caledonia | -19.3 | Macau | -18.3 | Yemen | -25.8 |
| 19 | Bulgaria | -19.1 | Bulgaria | -17.4 | Romania | -25.6 |
| 20 | Romania | -17.6 | Cape Verde | -17.2 | Saint Vincent and the Grenadines | -23.7 |

## Time series of total and sector-specific trade changes

The total trade losses are not uniform across sectors. Fig 3 shows the estimated total trade losses over time (Fig 3a) together with the trade losses for the 11 sector classification considered. The total trade losses are found to be between -7.0% and -9.6% (mean -8.3%), which is equal to around 206-286 MT in volume losses and up to 225-412 billion USD in trade value (uncertainty due to differences in total import and export losses and due to the volume to value conversion). The time series show (Fig 3a) a clear initial drop in trade in the first three months, after which trade partly recovers, followed by a second, more pronounced, drop in trade. In late August 2020, a sign of economic recovery is not yet visible.

Some supply-chains have been more resilient than others. The most resilient sectors are found to be Textiles and wearing apparel (-4.1%), Food and beverages (-5.8%), Other manufacturing (-6.0%) and Wood and paper (-6.3%). The times series of Textiles and wearing apparel and Other manufacturing show, however, a large drop in exports in the early stages of the pandemic, mainly associated with production in China and other Asian economies (e.g. Bangladesh, Malaysia), followed by a gradual recovery and a less steep second drop. The Wood and paper and Food and beverages sectors have been more stable throughout pandemic outbreak, as supply-chains were not significantly disrupted, and demand for products only gradually declined, followed by signs of a recovery at the end of August. The largest relative changes are found for the Fishing sector (-9.5%), Mining and quarrying (-9.0%), Manufacturing of electronics and machinery (-8.8%), and Manufacturing of transport equipment (-11.8%). The drop in fishing products peaked late in the pandemic with a clear recovery in July and August. The time series of the mining and quarrying sector shows a more complex picture with a sharp drop in the beginning of the pandemic, as demand for raw materials decreased in Asia, followed by a steep increase in trade to restock inventories, after which a total collapse of the market can be observed, mainly associated with reduced demand for oil. For the two manufacturing sectors, the large losses are the result of significant supply-chain disruptions that caused upstream production processes to halt due to a shortage in supplies [8]. In particular the Transport manufacturing industry, characterised by just-in-time logistic services and highly specialised production processes, experienced a gradual disruption throughout the first few months, after which trade declined more than 20% in May, June and July.

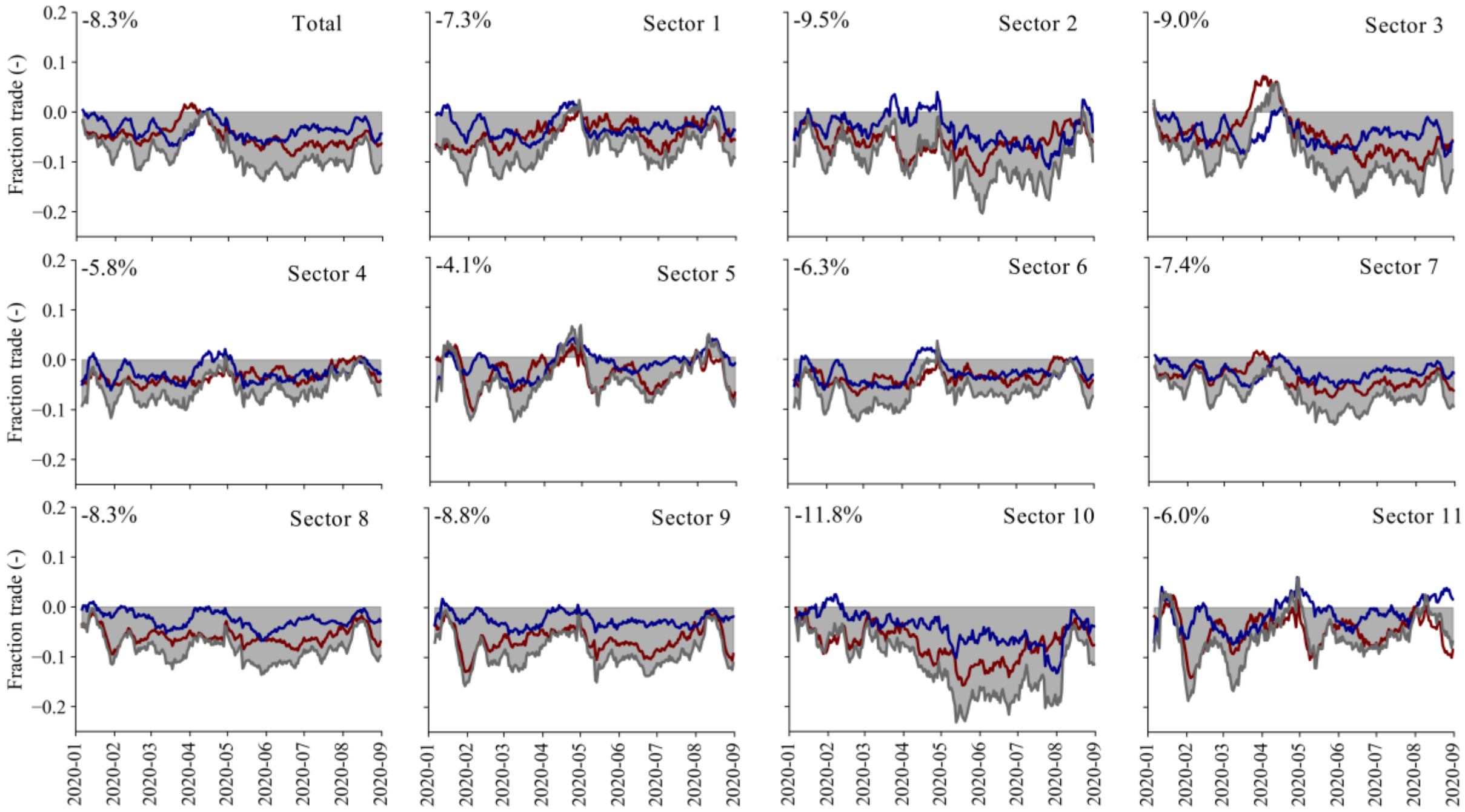

**Fig 3. Sector-specific losses over time.** The change in daily global total trade as a fraction of the average daily trade (over 2019). The dark blue line represent imports, the dark red line represent exports, whereas the grey line indicate total trade (import + exports). Sector 1: Agriculture; Sector 2: Fishing; Sector 3: Mining and quarrying; Sector 4: Food and beverages; Sector 5: Textiles and wearing apparel; Sector 6: Wood and paper; Sector 7: Petroleum, chemical and non-metallic mineral products; Sector 8: Metal products; Sector 9: Electrical and machinery; Sector 10: Transport equipment; Sector 11: Other manufacturing.

## Impact of large-scale containment measures

The results of the panel regression model are included in Table 1. After testing various model specifications (S1 Appendix), we find the most robust results for the model that includes daily control variables for the number of confirmed cases as a fraction of the population (Cases), demand reduction in trade dependent countries (Demand), the potential supply disruptions through changes in import that are used for exports (Supply), and potential other factors that are autocorrelated with the changes in export (Export lag).

The effect of the composite index on daily export change is strong, and statistically significant ($p < 0.01$), with a 10% increase of the index resulting in a -0.45% change in exports (Model 1).

The influence of containment measures on exports is mixed with some measures showing a negative impact while others showing a positive impact (Model 2). Negative impacts are found for school closing (C1, - 2.63%, significant at $p < 0.05$), workplace closure (C2, -4.76, significant at $p < 0.01$) and closing of public transport (C5, -3.59% significant at $p < 0.01$). Surpringly, a positive effect is found for stay at home requirements (C6, +2.74%, significant at $p < 0.10$), restrictions on internal movement (C7, +2.17%, significant at $p < 0.10$). Hence, implementing some policies has enabled countries to continue producing export goods and transport them to their ports, although the signal is generally not strong.

The results are potentially biased because of the fact that containment measures may affect domestic production in a lagged manner. In S2 Appendix, we perform additional robustness tests by introducing a 4/8/12/16 day lag of the containment policies. This increases the negative effect of school closures, which varies between -3.36% and -6.87% depending on the lag (all significant at $p < 0.01$), while it increases the positive effect of international travel bans for lags of 8 – 16 days (between +2.85% and 6.03%, significant at $p < 0.01$).

Additionally, we run a model (Model 3 and 4 that includes daily dummy variables to account for economy-wide factors that affect maritime exports) and a model (Model 5 and 6) which include only the days where the outbreak become significant in a country (which we define as having at least 50 confirmed cases). This coefficient of the composite index is stable across models (-4.53% versus -4.94% and -4.72%, Model 3 and 5). For the individual containment measures, the effect of school closures is larger in Model 4 (-2.63%, significant at $p < 0.05$) and Model 6 (-6.53%, significant at $p < 0.01$). The positive effect of testing is enhanced in Model 6 (+2.83%, significant at $p < 0.05$), whereas the positive effect of stay at home policies becomes larger and statistically significant in Model 4 and 6 (+2.74%, significant at $p < 0.10$; +6.53%, significant at $p < 0.01$). The results change little when performing additional robust checks by modifying the number of days of the lag of the export and adding week dummies (instead of day dummies) to the model (see S2 Appendix).

Across all model specifications, we find a consistent negative relationship between the overall stringency index and maritime exports, and business and public transport closures and maritime exports, whereas all other policies are either not significant, or not consistently significant across different model specifications.

**Table 3. The results of the various regression models.** The table shows the estimated beta coefficients and goodness of fit statistics for the six model specifications discussed. *p < 0.1, **p < 0.05, ***p < 0.01

| | Model1 | Model2 | Model3 | Model4 | Model5 | Model6 |
|---|---|---|---|---|---|---|
| Parameter | Beta | Beta | Beta | Beta | Beta | Beta |
| Composite | -4.527*** | | -4.939*** | | -4.717*** | |
| C1 | | -1.969 | | -2.630** | | -6.531*** |
| C2 | | -4.703*** | | -4.755*** | | -3.006* |
| C3 | | 0.557 | | 0.381 | | 1.298 |
| C4 | | -0.613 | | -0.822 | | -2.114 |
| C5 | | -3.588*** | | -3.500*** | | -5.429*** |
| C6 | | 2.049 | | 2.740* | | 6.525*** |
| C7 | | 2.173** | | 2.191* | | 1.377 |
| C8 | | 1.825 | | 1.995 | | 3.771 |
| H2 | | 0.890 | | 0.728 | | 2.833** |
| Demand | -0.015 | -0.033 | -0.016 | -0.014 | 0.005 | -0.022 |
| Cases | -0.129** | -0.113* | -0.130* | -0.124* | -0.393*** | -0.427*** |
| Supply | 0.018*** | 0.018*** | 0.018*** | 0.018*** | -0.003 | -0.002 |
| Export lag | 0.350*** | 0.349*** | 0.351*** | 0.350*** | 0.402*** | 0.399*** |
| R2 | 0.320 | 0.321 | 0.329 | 0.330 | 0.359 | 0.361 |
| R2-adjusted | 0.316 | 0.316 | 0.318 | 0.319 | 0.353 | 0.355 |
| F-statistic | 73.67 | 69.23 | 29.84 | 29.20 | 64.74 | 61.18 |

# Discussion and conclusion

We present a near-global analysis of maritime trade indicators based on empirical vessel tracking data. We illustrate how the implementation of large-scale containment measures have resulted in large trade losses, and hence domestic output, with a strong geographical and sectoral heterogeneity.

Our estimate of a 4.4% reduction in global ports calls for the first eight months of 2020 is lower than the 8.7% predicted by UNCTAD for the first six months [14]. The main reason for this difference is associated with the inclusion of different vessel types. Whereas we include only the main trade-carrying vessels, the UNCTAD analysis also included passengers vessels (66% of total port calls), which have seen the largest drop in port calls (-17% for passenger vessels). Moreover, the sector-level trends we found are in line with the sector-level impacts (based observed trade data of China, the European Union and the United States) for the first quarter (Q1) of 2020 as presented in the UNCTAD analysis [14], that stated that in particular the automotive industry (-8%), machinery (-8%), office machinery (-8%) and textiles and apparel (-11%) are particularly hit. Our analysis, which differs by only including maritime trade (instead of total trade) and having a global scale (instead of three countries), found the average losses in Q1 for the textiles and apparel (Sector 5), electrical equipment and machinery manufacturing (Sector 9), transport equipment (Sector 10) and other manufacturing (Sector 11) to be respectively 6.2%, 9.2%, 6.5% and 8.4%. The trade losses we estimate (225-412 billion USD for the first eight months) are considerably lower than reported in the modelling framework of Lenzen et al. [7], who estimated the trade losses for the first five months of 2020 to be 536 billion USD. Again, part of this difference is due to the coverage of countries (they provide a full global analysis) and modes of transport (all modes compared to maritime only). Still, input-output based analysis, as done in Lenzen et al. [7], often fail to consider adaptative behaviour in the global economic system, which can dampen economic impacts [33]. The -8.6% trade losses we found are also lower than the 26.8%, and up to 31%, value-added losses (although not being directly comparable) reported in the global pandemic scenarios of Guan et al. [6]. We expect the percentage change in trade losses to increase going further in 2020, as daily trade losses are above 10% at the end of August. This is in line with projections of the UNCTAD [14] and the WTO [10] that estimate a further contraction of merchandise trade in the second half of 2020.

The results of the econometric model provide an alternative view to the studies that have evaluated the effect of large-scaled containment measures on the spread of the virus [1,2,4,34]. We find clear evidence of the negative impacts of large-scale containment measures on changes in daily exports, with a 10% increase in the overall stringency value resulting in a 0.45%-0.50% decrease in daily maritime exports. In particular, introducing required closing of all public transport and all-but-essential businesses resulted in large maritime export losses of up to 5.4-6.5% of daily export. This is in agreement with Deb et al. [5], who used nitrogen emissions as an indicator of industrial output and found that workplace closures had the largest influence on the drop in emissions. Results should, however, be interpreted with caution, as many factors could potentially influence these causal relationships. For instance, temporal increases in maritime transport during some periods of the pandemic could be driven by the large increase in trade of medical supplies (e.g. PPE) and mode substitution from air to maritime [35], irrespective if policies were imposed during these periods. Therefore, testing alternative economic indicators, such as data on retail sales, mobility, flight, energy consumption and nitrogen emissions, as done in Deb et al. [5] can help support these findings. Overall, our analysis of the economic implications of introducing containment policies into society can help evaluating the cost-benefit of the different containments measures, which may help governments construct effective portfolios of policies as many countries enter a second-wave of COVID-19 cases [36]. Future work can refine our estimates by adding more trade data when it becomes available, including extending the analysis to indicators of air, road and rail transport. Moreover, the empirical estimates derived here can be used to constrain and validate macro-economic impact models in order to improve the quantification of the total losses to industrial output as the pandemic unfolds.

Overall, real-time indicators of economic activity, such as maritime trade, can help identify trends in supply-chains networks and support governments and international organisations in their economic recovery efforts by allocating funds to the hardest hit economies and sectors.

## Acknowledgements

All derived datasets used for this analysis are made publicly available at Zenodo: 10.5281/zenodo.4146993 and will be published upon acceptance. The policy indicators are obtained from the Oxford Coronavirus Government Response Tracker (https://www.bsg.ox.ac.uk/research/research-projects/coronavirus-government-response-tracker).

The authors would like to thank the United Nations Statistical Division and the UN Global Working Group on Big Data for Official Statistics, in particular Markie Muryawan and Ronald Jansen, for providing the AIS data.

## Supporting information

**S1 Appendix: Methodology maritime trade estimates**

**S2 Appendix: Econometric model**

**S1 Table. The top 20 largest negative maritime trade losses on a country-level.** The total trade, imports and exports losses expressed in million tonnes (MT). The losses cover the period Jan-Aug 2020 compared to Jan-Aug 2019.

**S1 Table. The top 20 largest negative maritime trade losses on a country-level.** The total trade losses and imports and export losses expressed in million tonnes (MT). The losses cover the period Jan-Aug 2020 compared to Jan-Aug 2019.

| | Total trade | | Imports | | Exports | |
|---|---|---|---|---|---|---|
| Rank | Country | Change (MT) | Country | Change (MT) | Country | Change (MT) |
| 1 | China | -342.4 | China | -217.8 | China | -124.6 |
| 2 | Saudi Arabia | -98.3 | United States | -32.4 | Saudi Arabia | -99.1 |
| 3 | United States | -91.3 | Netherlands | -28.6 | Australia | -60.7 |
| 4 | Australia | -65.1 | Japan | -17.7 | United States | -58.9 |
| 5 | United Arab Emirates | -47.3 | India | -14.2 | Brazil | -39.5 |
| 6 | Netherlands | -46.2 | United Arab Emirates | -9.8 | United Arab Emirates | -37.6 |
| 7 | Brazil | -44.2 | Italy | -8.8 | Japan | -18.8 |
| 8 | Japan | -36.5 | Mexico | -8.5 | Russia | -18.7 |
| 9 | Canada | -21.9 | Great Britain | -8.3 | Netherlands | -17.6 |
| 10 | Russia | -20.3 | Belgium | -7.2 | Columbia | -16.2 |
| 11 | Columbia | -17.1 | Kuwait | -7.1 | Canada | -15.5 |
| 12 | Peru | -16.9 | Canada | -6.3 | Peru | -13.0 |
| 13 | Italy | -16.4 | Brazil | -4.7 | Malaysia | -9.7 |
| 14 | Malaysia | -13.0 | South-Korea | -4.6 | South-Africa | -8.1 |
| 15 | South-Korea | -12.5 | Australia | -4.4 | South-Korea | -8.0 |
| 16 | Belgium | -11.9 | France | -4.3 | Libya | -7.9 |
| 17 | Mexico | -11.9 | Iran | -4.0 | Italy | -7.7 |
| 18 | Kuwait | -10.8 | Peru | -3.8 | Ukraine | -6.9 |
| 19 | Great Britain | -9.1 | Egypt | -3.8 | Singapore | -6.5 |
| 20 | Thailand | -8.6 | Thailand | -3.5 | Indonesia | -6.0 |

**S2 Appendix: econometric model**

***Model description and formulation***

To estimate the effect of various large-scale containment measures on daily maritime exports, we apply reduced-form econometrics [1] using a fixed effect panel regression model. This type of model has been used previously to estimate the effect of large-scale containment measures on the spread of COVID-19 [2,3] and economic output (using nitrogen emissions as an indicator) [3]. We follow a similar implementation of the model as done in previous studies, although control more specifically for the potential supply and demand shock compared to for instance Deb et al. [3].

For every country (i), we derive country-wide daily (t) indicators of the change in maritime exports ($\Delta E_{i,t}$) by aggregating the daily time series of estimated export data on a port-level. We use exports as an indicator of industry output as it better reflects the status of the economy compared to changes in imports, which are likely influenced by supply shortages (due to reduced output in trade-dependent countries) and reduced demand for products (due to imposed lockdowns). First, we detrend (using a linear regression) the 2019 time series in order to filter out a structural increase or decrease in maritime exports. The trend is only removed if a statistically significant trend is observed ($p < 0.05$). Second, we smooth the data using a 10 day moving average, which is necessary to reduce the noise for countries with variable daily export estimates (e.g. smaller economies). Third, we compare the 2020 time series (Jan – Aug) to the time series for 2019 and estimate the percentage change deviation from the average daily exports (based on 2019 daily data). We remove countries where the daily exports are zero for at least one day, as this will likely bias the results (e.g. some small islands only export products on certain days depending on the arrival of maritime vessels).

For every country, we retrieve information about the government policy responses from the Oxford COVID-19 Government Response Tracker (OxCGRT) [4]. We obtained information on nine measures that potentially affect business operations: C1 - School closing; C2 - Workplace closing; C3 - cancel public events; C4 - Restrictions on gatherings; C5 - Close public transport; C6 - Stay at home; C7 - Restrictions on internal movement; C8 - International travel controls; H2 - Testing policy. These measures all have different ordinal scales depending on the different levels of responses (e.g. restrictions internal movements has two levels, whereas international travels restrictions has four levels). We normalise all policies to a 0 to 1 range, with 0 implying no measure implemented and 1 referring to the maximum severity of the measure. This provides us with a set of daily policy responses (p) per country ($c_{p,i,t}$). We also derive an overall Stringency metric that is the sum of all policies (C1-C8) normalised to a 0 to 1 scale ($S_{i,t}$).

We control for several factors ($X_{i,t}$), including (1) the daily number of confirmed cases, (2) the supply-shock, and (3) the demand-shock. These factor can cause variations in the export change across countries. The daily number of confirmed cases are obtained from OxCGRT [4], which we divide over the country's population in order to estimate the fraction of the population that is positively tested on a particular day. The estimate the supply-shock, we derive the daily, sector-specific, changes in imports that are used to produce export products. To estimate the percentage of sector-specific imports that are used for exports, we create a sector-specific vertical specialisation coefficient per country, as proposed by Hummels et al. [5], using the 2015 EORA multi-regional input-output tables [6]. This coefficient reflect the dollar value increase in imports for every dollar increase in exports in a country, including all industry interdependencies. By multiplying this coefficient with the sector-specific daily data per country, we estimate the daily time series of supply of goods that are used to produce exports. We again derive daily percentage changes in the supply (similar as for the export time series, see above). We implement this supply time series with a 7 day lag to account for lags in the use of input products and to account for the potential buffering capacity of inventories. The daily time series of the demand shock is derived by estimating the weighted average stringency value in trade-dependent countries, assuming that countries that have higher stringency values demand less products. The stringency values are derived from OxCGRT [4] and range between 0 and 100. We extract data from the 2018 (latest year available) BACI harmonized trade database [7] and use this to estimate the daily demand shock by multiplying the stringency value of the importing country with the fraction the bilateral trade flow (between exporting and importing country) contributes to the total exports of exporting country (weight). For every export country, we sum over the total number of trade relationships to end up with a weighted mean stringency value. In this way, we account for the fact that exports can reduce before a country implemented containment policies (as was the case in the beginning of the pandemic). We hereby assume that a demand shock happens instantaneously without any lag.
At last, we add a three day lag of the export change ($\Delta E_{i,t-\Delta t}$) itself to the model in order to effectively control for the normal dynamics in daily exports and other endogenous factors that are likely to be serially correlated with daily exports.

In summary, we can express the daily change in exports in 2020 compared to 2019 as:

$$\Delta E_{i,t} = \mu_i + c_{p,i,t} + X_{i,t} + \Delta E_{i,t-\Delta t} + \epsilon_{i,t}$$

with $\mu_i$ the country-specific fixed effect to account for time-invariant country characteristics. We also tested implementing the policies as separate dummy variables per ordinal scale, in order to capture non-linear effects, which gives similar results, although it becomes harder to estimate the effect of the individual policies.
Alternatively, we can replace the individual policy interventions with the overall Stringency estimate:

$$\Delta E_{i,t} = \mu_i + S_{i,t} + X_{i,t} + \Delta E_{i,t-\Delta t} + \epsilon_{i,t}$$

The results of these models are included in Table 1 in the main article.

***Robustness checks***

In order to account for the fact that the policies might effect exports in a lagged way, we implement the individual policies in a lagged manner ($c_{p,i,t-\Delta t}$) with $\Delta t$ = 4/8/12/16 days:

$$\Delta E_{i,t} = \mu_i + c_{p,i,t-\Delta t} + X_{i,t} + \Delta E_{i,t-\Delta t} + \epsilon_{i,t}$$

The results are shown in S2 Table 1.

**S2 Table 1: Regression results for the model specifications that include additional time lags in introducing the individual containment measures. *p < 0.1, **p < 0.05, ***p < 0.01**

| | **Base** | **4day lag** | **8day lag** | **12day lag** | **16day lag** |
|---|---|---|---|---|---|
| Parameter | Beta | Beta | Beta | Beta | Beta |
| C1 | -1.9685 | -3.3580*** | -6.1903*** | -6.8629*** | -4.8998*** |
| C2 | -4.7032*** | -3.5833*** | -2.9778** | -2.4064* | -2.3936* |
| C3 | 0.5571 | 1.3945 | 1.8149 | -0.3604 | -1.6499 |
| C4 | -0.6133 | -0.2415 | 0.2850 | -0.1466 | -1.5849 |
| C5 | -3.5877*** | -2.9857*** | -3.4684*** | -2.6925*** | -1.5073 |
| C6 | 2.0491 | 1.6623 | 4.5472*** | 5.6175*** | 4.1332*** |
| C7 | 2.1731** | 0.2912 | 0.0685 | 0.5809 | 2.2353** |
| C8 | 1.8246 | 1.5980 | 2.8517** | 4.3879*** | 6.0277*** |
| H2 | 0.8896 | 0.5020 | 0.3387 | 0.4305 | -0.6363 |
| Demand | -0.0326 | -0.0151 | -0.0288* | -0.0312** | -0.0527*** |
| Cases | -0.1134* | -0.0769 | -0.0602 | -0.0342 | -0.0495 |
| Supply | 0.0183*** | 0.0176*** | 0.0177*** | 0.0178*** | 0.0181*** |
| Export lag | 0.3491*** | 0.3491*** | 0.3488*** | 0.3481*** | 0.3481*** |
| R2 | 0.321 | 0.321 | 0.322 | 0.322 | 0.322 |
| R2-adjusted | 0.316 | 0.316 | 0.317 | 0.318 | 0.317 |
| F-statistic | 69.23 | 69.25 | 69.48 | 69.65 | 69.53 |

Moreover, we include a dummy for the week/day ($\tau_t$) to account for changes in the global economy that effect exports across countries:

$$\Delta E_{i,t} = \mu_i + \tau_t + c_{p,i,t} + X_{i,t} + \Delta E_{i,t-\Delta t} + \epsilon_{i,t}$$

The results for the week and day dummy variables are included in S2 Table 2.

**S2 Table 2: Regression results for the model specifications that include additional time dummies. *p < 0.1, **p < 0.05, ***p < 0.01**

| | **Base** | | **Week dummy** | | **Day dummy** | |
|---|---|---|---|---|---|---|
| Parameter | Beta | 95% CI | Beta | 95% CI | Beta | 95% CI |
| C1 | -1.9685 | [-4.369, 0.432] | -2.6353** | [-5.078, -0.192] | -2.6299** | [-5.079, -0.180] |
| C2 | -4.7032*** | [-7.327, -2.080] | -4.6779*** | [-7.366, -1.990] | -4.7550*** | [-7.451, -2.059] |
| C3 | 0.5571 | [-2.085, 3.200] | 0.0589 | [-2.680, 2.798] | 0.3805 | [-2.375, 3.136] |
| C4 | -0.6133 | [-3.049, 1.823] | -0.8523 | [-3.346. 1.641] | -0.8215 | [-3.320, 1.677] |
| C5 | -3.5877*** | [-5.692, -1.483] | -3.4802*** | [-5.593, -1.367] | 3.5003*** | [-5.616, -1.385] |
| C6 | 2.0491 | [-0.958, 5.056] | 2.7083* | [-0.332, 5.748] | 2.7399* | [-0.305, 5.785] |
| C7 | 2.1731** | [-0.002, 4.348] | 2.2822** | [0.097, 4.467] | 2.1912* | [0.002, 4.380] |
| C8 | 1.8246 | [-0.597, 4.246] | 1.7982 | [-0.696, 4.292] | 1.9948 | [-0.508, 4.498] |
| H2 | 0.8896 | [-1.434, 3.213] | 0.4548 | [-2.179, 3.089] | 0.7284 | [-0.508, 4.498] |
| Demand | -0.0326 | [-0.078, 0.012] | -0.0325 | [-0.096, 0.031] | -0.0139 | [-0.078, 0.012] |
| Cases | -0.1134* | [-0.235, 0.008] | -0.1080* | [-0.246, 0.030] | -0.1240* | [-0.263, 0.015] |
| Supply | 0.0183*** | [0.008, 0.029] | 0.0184*** | [0.008, 0.029] | 0.0176*** | [0.007, 0.028] |
| Export lag | 0.3491*** | [0.340, 0.358] | 0.3478*** | [0.339, 0.357] | 0.3496*** | [0.340, 0.359] |
| R2 | 0.321 | | 0.325 | | 0.330 | |
| R2-adjusted | 0.316 | | 0.320 | | 0.319 | |
| F-statistic | 69.23 | | 58.22 | | 29.20 | |